\documentstyle[aps,prb,preprint]{revtex}
\begin{document}
\draft
\title{ Coupled-barrier diffusion: the case of oxygen in silicon  }
\author{Madhavan Ramamoorthy and Sokrates T. Pantelides}
\address{Department of Physics and Astronomy, Vanderbilt University,
Nashville, TN 37209 }
\date{\today}
\maketitle
\begin{abstract}

Oxygen migration in silicon corresponds to an apparently simple jump between
neighboring bridge sites. Yet, extensive theoretical calculations have so
far produced conflicting results and have failed to provide a satisfactory
account of the observed $2.5$ eV activation energy. We report a comprehensive
set of first-principles calculations that demonstrate that the seemingly
simple oxygen jump is actually a complex process involving coupled barriers
and can be properly described quantitatively in terms of an energy
hypersurface  with a ``saddle ridge'' and an activation energy of
$\sim 2.5 $ eV. Earlier calculations correspond to different points or lines
on this hypersurface.

\end{abstract}

\vspace{0.1in}


\narrowtext

 Oxygen in silicon has long been known to occupy a bridge position between
neighboring Si atoms, with an Si-O-Si configuration similar to those
in SiO$_2$.\cite{corb64,ynd87.2}
Its diffusion, measured to have an activation
energy of $2.5$ eV,\cite{mikk82}
is generally believed to consist of simple jumps between
neighboring bridge positions on the (110) plane
defined by the corresponding Si-Si bonds (Fig. 1). In terms of
the angle $\theta_{\rm O}$ defined  in Fig. 1, the
midpoint of the jump is at $\theta_{\rm O} = 90^{\circ}$.

Most calculations to date\cite{snyd86,kell92,sai88,osh90,need91}
assumed such a simple adiabatic jump, with  reflection symmetry about the
vertical axis shown in Fig. 1. Thus, the saddle point was assumed to have the
O atom at $\theta_{\rm O} = 90^{\circ}$ and the central Si atom at
$\theta_{\rm Si} = 90^{\circ}$.
The remaining degrees of freedom
and the positions of the other  Si atoms were
determined by total-energy minimization.
The resulting total energy, measured from the energy of
the equilibrium configuration, represents the adiabatic activation energy for
diffusion. Some authors\cite{snyd86,kell92}
reported activation energies around $2.5$ eV, while
others\cite{sai88,osh90,need91} reported smaller values ranging from
$1.2$ to $2.0$ eV.

In Ref. 8,  Needels et al. found a value of $1.8$ eV and
attributed the discrepancy with experiment to
dynamical phenomena, i.e., the neighboring Si atoms do not relax fully
along the O trajectory. They reported model dynamical calculations
for a ``generic'' non-adiabatic path in
which the O atom was given an initial ``kick'', i.e.,
an initial velocity
corresponding to  a kinetic energy of $2.0$, $2.3$ or  $2.7$ eV.
They found that when the the kick energy was  $ < 2.5$ eV,
the O atom went past
the saddle-point but then returned to the original bridge position.
When the kick was $> 2.5$ eV the O atom migrated to the next bridge
site. They concluded that their results suggested that dynamical effects
are important in O migration, but did not constitute definitive
evidence.

In a recent paper, Jiang and Brown\cite{jiang95} (JB)
sought to resolve the issue by exploring the entire migration path.
They performed total-energy minimizations by stepping the oxygen atom from one
bridge site to the next. They found that the  total
energy attains a value of only $\sim 1.2$ eV at $\theta_{\rm O} = 90^{\circ}$,
but then keeps rising to a maximum value (saddle point) of $\sim 2.5$ eV at
$\theta_{\rm O} = 113^{\circ}$. In addition, they computed the diffusion
constant and found it to agree very well with experiment over 12 decades.
They concluded that the saddle point of O migration is past the midpoint of
the path and that their results account for all the experimental data.

At first glance, JB's results account nicely for the experimental data
without the need to invoke dynamical effects. Nevertheless,
the pronounced asymmetry of JB's
total-energy profile about $\theta_{\rm O} = 90^{\circ}$ raises a serious
question: if the global minimum of the total-energy was indeed obtained at
each point of the O path, the energy profile would be symmetric about
$\theta_{\rm O} = 90^{\circ}$. Clearly, JB's minimization procedure
yielded a secondary minimum for each $\theta_{\rm O} > 90^{\circ}$, not the
global minimum.
The energy at the global minimum for $90^{\circ} + \alpha$
is by symmetry equal to that at $90^{\circ} - \alpha$.
If an energy profile were constructed using global minima along the
entire path, it would have a maximum of
only $1.2$ eV at $\theta_{\rm O} = 90^{\circ}$.
This value would be in poor agreement with experiment.
We conclude that there is still no satisfactory account of the
observed $2.5$ eV activation energy, or of the mutually conflicting
theoretical results published so far on oxygen diffusion in silicon.

In this paper, we report a series of first-principles
total-energy calculations which show
that the process of O migration is far more complex than has been recognized
so far, but is still adiabatic.
During the migration process, {\it both} the O atom and the central Si atom
perform jumps and face large barriers.
As a result, a quantitative description of the process
requires a calculation of the total-energy hypersurface as a function of
the positions of both atoms.
We will show slices of this hypersurface that reveal a ``saddle ridge''
in multidimensional space. The migration process is adiabatic and occurs along
a multiplicity of paths over this ridge with a predominant barrier of
$\sim 2.5$ eV. Finally, we find that the results of earlier authors correspond
to different points or lines on the hypersurface.

We performed calculations using density functional theory and the local-density
approximation for exchange and correlation, using the form for the
exchange-correlation potential given by  Ceperley and Alder.\cite{cep80}
The ultra-soft pseudopotentials of Vanderbilt\cite{vand90} were used for
Si and O.
These  pseudopotentials have been thoroughly tested in several extensive
investigations.\cite{king94,feng94,ramamoor94.2}
The  calculations employed a plane wave basis set and
converged results were obtained with an energy cutoff of 25 Ry.
A  bcc supercell with
32 Si atoms and one O atom was used.
Each structure was relaxed until the force on each atom
was less than $0.5$eV/$\AA$.
All calculations were first done
with one special k-point at $(0.5, 0.5, 0.5)$ in the irreducible Brillouin
zone.\cite{chadi73} The key calculations were repeated with 2 k-points
at  $(0.75, 0.25, 0.25)$ and $(0.25, 0.25, 0.25)$.\cite{chadi73}
The energy differences changed
at most by about $0.2$ eV, with all the qualitative results obtained
with one k-point being unchanged.
Hence, the results with one k-point were taken
to be converged with respect to k-point sampling, and used in all
the figures in this paper.

Our results for the equilibrium  configuration of O, shown schematically
in Fig. 1,  are in agreement with
earlier work.\cite{ynd87.2} We find a very flat minimum at
$\theta_{\rm O} \sim 55^{\circ}$. The Si-O bond length is $1.6$ \AA, the Si-Si
length is $3.2$ \AA \hspace{0.01in}
(compare with the value of $2.35$ \AA \hspace{0.01in} in bulk silicon)
and the Si-O-Si bond angle is $\sim 150^{\circ}$.
The angle $\theta_{\rm Si}$ is also $\sim 150^{\circ}$.
For our purposes here, the key point is that the central Si atom is
well to the right of the vertical symmetry axis (see Fig. 1).
As the O migrates from the left bridge position to
the one on the right, the central Si needs to move from the right to the left,
specifically from $\theta_{\rm Si} \sim 150^{\circ}$ to
$\theta_{\rm Si} \sim 30^{\circ}$. We will see below that {\it this swing of
the central Si atom controls the dynamics of the oxygen migration because
the Si atom has to overcome a barrier}.

We demonstrate this key result in Fig. 2 where we plot the total energy of the
system as a function of $\theta_{\rm Si}$  when the O atom at
$\theta_{\rm O} = 90^{\circ}$.\cite{foot_1}
For each $\theta_{\rm Si}$, the total energy was minimized
with respect to ${\rm R}_{\rm O}$, ${\rm R}_{\rm Si}$ (see Fig. 1)
and the positions of the other Si atoms.
We see that it costs only $0.6$ eV to place the O atom at the midpoint if the
central Si atom is allowed to relax freely either to the left or to the
right!\cite{foot_2}
As we saw earlier, as the O atom moves from the left bridge position to the
one on the right, the central Si atom needs to swing from the right side to
the left side. Figure 2 shows that, with
$\theta_{\rm O} = 90^{\circ}$, the total barrier is $2.2$ eV. This barrier
corresponds to the two atoms crossing the midpoints of their respective
paths at the same time.
It could be argued that this configuration  constitutes the
saddle point, as was assumed in several previous
investigations.\cite{snyd86,kell92,sai88,osh90,need91}
The total barrier of $\sim 2.2$ eV is indeed in good
agreement with the experimental value. This simple result, however, belies
an enormous complexity which we unravel below. The basis of this complexity
is that the O atom and the central Si atom need not pass through the
midpoints of their paths at the same time.

The above analysis  makes it clear that O migration needs
to be described in at least a
two-dimensional space defined by $\theta_{\rm O}$ and $\theta_{\rm Si}$
because the central Si atom also must climb a barrier.
This barrier, however, is not simple, but, as shown in Fig. 2,
has a cusp at $\theta_{\rm Si} =
90^{\circ}$, indicative of a Jahn-Teller-like instability with
two symmetric total-energy manifolds.
These two manifolds correspond to the central Si atom being to the left or
the right of the symmetry axis, being bonded to the respective Si atom
on the left or the right.
For values of $\theta_{\rm O}$ other than $90^{\circ}$, the two manifolds are
not symmetric.
In Fig. 3 we trace the evolution of the two total-energy manifolds for a
sequence of $\theta_{\rm O}$ values starting with the O atom
near its equilibrium bridge position on the left of the vertical
symmetry axis (bottom panel).
The central Si atom is on the right side of the axis
(the lower-energy manifold). As the O atom progresses along its path
(higher panels in Fig. 3) the central Si atom stays in the right manifold.
At $\theta_{\rm O} = 90^{\circ}$, the two manifolds cross
and the central Si
atom can switch manifolds and swing over to the left of the axis,
so that both the O atom and the central Si atom can head for their
final destinations. The total barrier for this process is $2.2$ eV.

 There are additional possibilities, however. The
O atom may overshoot the midpoint of its jump without the central Si atom
swinging over. The relevant total-energy manifolds are shown in the upper
panels of Fig. 3. The central Si atom is  now in the high-energy manifold,
stuck on the ``wrong'' side of the vertical axis.
Even though the manifolds
do not cross, the central Si atom is stable in the higher-energy manifold
only up to a certain value of $\theta_{\rm Si}$, marked by the solid arrows.
At those points, the calculations show that the central Si atom collapses
to the lower-energy manifold, i.e. swings over to the left side of the
axis. No matter how much the O atom overshoots (i.e. any
of the top four panels), the total energy needed for the central Si atom
to swing over is of order $2.3-2.7$ eV.
In other words, the O
atom and the central Si atom {\it need not move in concert and
be at the midpoints of their respective paths at the same time}.
They can move independently and
still face a net barrier of $\sim 2.5$ eV.

The multiplicity of paths is best illustrated with a two-dimensional plot
of the total energy as a function of $\theta_{\rm O}$ and $\theta_{\rm Si}$.
For clarity, we show only one of the two manifolds at each pair
($\theta_{\rm O}$, $\theta_{\rm Si}$), namely the one corresponding to
coupled adiabatic migration. The surface was obtained by interpolating
through a sizable number of calculated points.
Note the  flat regions corresponding to the two equilibrium configurations
and the steep drops that correspond
to the regions of the solid arrows in Fig. 3. We see that there is a
``saddle ridge'' with a net energy of $\sim 2.5$ eV over a considerable range.
At the high symmetry point on the ridge ($\theta_{\rm O} = 90^{\circ}$,
$\theta_{\rm Si} = 90^{\circ}$), the total energy is somewhat lower,
$\sim 2.2$ eV, but this smaller value corresponds to a small fraction of all
possible  migration paths over the ridge (the resolution of the figure
is limited by the complexity of the surface near the ridge). There
are two classes of paths: those in which the O atom overshoots the midpoint
of its path with the central Si atom trailing, and those in which the
central Si atom goes over the midpoint of its path first with the O atom
trailing. Along all these paths the distance between the O atom and the central
Si atom is $\sim 1.7$ \hspace{0.01 in} \AA.
Thus, the O-Si bond acts like a pogo stick that faces
a net barrier of $\sim 2.5$ eV no matter how it turns
as it attempts to change its tilt from the left to the right.

The collapse from one manifold to the other indicated by the solid arrows in
Fig. 3 (steep drops in Fig. 4) was intriguing enough to merit further
investigation. The plots in Fig. 3 and Fig. 4 were constructed by picking
$\theta_{\rm O}$ and $\theta_{\rm Si}$ and then letting both the O atom
and the central Si atom move {\it radially} until the energy was
minimized. It turns out that the two manifolds shown in Fig. 3 correspond to
two fairly distinct regions of R$_{\rm Si}$ values. We explored R$_{\rm Si}$
values between these two regions and found that for each $\theta_{\rm Si}$,
the energy as a function of R$_{\rm Si}$ has two minima with a barrier
that prohibits the central Si atom's  motion from one minimum to another,
corresponding to a switch between the two energy manifolds.
At the critical $\theta_{\rm Si}$ value (solid arrow in Fig. 3), this barrier
vanishes and the collapse occurs. The evolution of this radial barrier is
also quite intriguing and will be discussed further in a longer article.
In fact a complete description of O migration requires the total energy
as a function of four coordinates on the (110) plane:
($\theta_{\rm O}$, $\theta_{\rm Si}$, R$_{\rm O}$, R$_{\rm Si}$).
Figures 3 and 4 represent slices through this hypersurface.

We now turn to examine the earlier theoretical work in the light of the
present work. The major point is that all earlier investigators did not
recognize the important role of the central Si atom in the migration
process. Nevertheless, we can map their results on Fig. 4.
References 4-8 assumed that the saddle point is at $\theta_{\rm O} =
\theta_{\rm Si} = 90^{\circ}$, shown as a dot in Fig. 4.
Our value for this point lies in the
middle of the range of earlier theoretical results ($1.8-2.5$ eV).\cite{foot_3}
This  spread in part reflects the fact that calculations involving oxygen are
computationally extremely demanding.
The important point to note is that
in Fig. 4, the paths that contain the point
$\theta_{\rm O} = \theta_{\rm Si} = 90^{\circ}$ constitute
a small portion of the phase space of  all paths going over the ridge.
Most of the ridge is somewhat higher, $\sim 2.5$ eV, close to the observed
activation energy.

Figure 4 also clarifies the ``kick'' simulations of
Needels et al.\cite{need91} and JB's calculations.
Both calculations correspond to the path shown by the hand-drawn
line in Fig. 4. Needels et al. gave a high kinetic energy to the O atom
whereas JB stepped the O atom gradually. In both cases, the O atom crossed
the midpoint of its path before the central Si atom did. In JB's case,
the central Si atom crossed over when $\theta_{\rm O} \sim 115^{\circ}$.
In the case of Needels et al., for ``kick'' energies $< 2.5$ eV, the O atom
had to turn back because the central Si atom still faced a barrier and
could not cross over. In hindsight, one should have given a kick to the
central Si atom. In any case, this is not a very likely path because in
reality both the O atom and the central Si atom are vibrating about
their equilibrium
positions, attempting to overcome their respective barriers. As the ridge has
roughly a constant height over a considerable range,
any point at an energy of $\sim 2.5$
eV may be fairly representative of the entire ridge. This might explain
the good agreement obtained by JB for the diffusion constant with experiment.

Finally, Needels at al.\cite{need91} noticed the formation of a metastable
configuration at the endpoint of the O path, when the kick was $2.7$ eV.
The present work shows that this configuration occurs in the
right-hand manifold shown in
the upper panels of Fig. 3. The metastable configuration is created
{\it during} the migration process for all paths that do not cross
$\theta_{\rm O} = \theta_{\rm Si} = 90^{\circ}$.
Contrary to the findings of Needels et al.,\cite{need91} this configuration
collapses to the equilibrium configuration
at the end of the migration process.

In summary, we have shown that the total-energy variation during O
migration is a very complex hypersurface in a multidimensional space. We have
shown slices of this hypersurface along some relevant coordinates,
revealing seemingly disconnected manifolds.
We believe that the calculations we have done so far have
captured essentially all of the very complex physics of the seemingly simple
O jump in bulk Si. Our results suggest that such complexity is
likely to be present whenever migration of an impurity involves
bond-breaking and rebonding with different host atoms.

\acknowledgments

This work was supported in part by the Office of Naval Research Grant No:
N00014-95-1-0906. Supercomputing time was provided by the
Pittsburgh Supercomputing Center under Grant No: DMR950001P.


\begin{figure}
\caption{ The geometry of O migration in a (110) plane. The solid dots
are the nominal positions  of Si atoms in the perfect crystal. The open
circles show the positions of the Si and O atoms  in the equilibrium
configuration.}
\label{fig:1}
\end{figure}

\begin{figure}
\caption{ The total-energy variation of the system,
as a function of $\theta_{\rm Si}$  when the O atom is at
$\theta_{\rm O} = 90^{\circ}$.
The zero of the energy in this
and in subsequent plots is taken as that of the equilibrium configuration.}
\label{fig:2}
\end{figure}

\begin{figure}
\caption{ A series of plots indicating the variation of the total energy
as a function of $\theta_{\rm Si}$ for several  values of
$\theta_{\rm O}$.}
\label{fig:3}
\end{figure}

\begin{figure}
\caption{ The total-energy variation as a function of $\theta_{\rm O}$ and
$\theta_{\rm Si}$. The central dot and the hand drawn path (in the direction
of the arrowhead) are discussed in the text.}
\label{fig:4}
\end{figure}

\narrowtext


\begin{references}
\bibitem{foot_1} In  all the calculations reported in this paper, we
confined the motion of the O atom and the central Si atom to the (110) plane.
Our test calculations showed that, in agreement with Ref. 9, the energy
rises if either atom is allowed to go off the (110) plane.
\bibitem{foot_2} In the symmetrical configuration $\theta_{\rm O} =
\theta_{\rm Si} = 90^{\circ}$, O is forced to have
three  bonds,
with each Si-O-Si bond angle about $85^{\circ}$.
In the  configurations corresponding to the minima in Fig 2,
there are only two Si-O  bonds with lengths $\sim 1.7$ \AA \hspace{0.01in}
and  the  magnitude of the Si-O-Si bond angle equal to $127^{\circ}$,
significantly closer to  equilibrium values.
\bibitem{foot_3} The value $1.2$ eV of Ref. 6 was replaced by a more accurate
value of $2.0$ eV by the same authors in Ref. 7.
\end{references}
\end{document}